\documentclass[prb,twocolumn,showpacs]{revtex4}
\begin{document}
\title{Doping Dependence of Vortex Regimes in Y$_{1-x}$Pr$_{x}$Ba$_{2}$Cu$_{3}$O$_{7-\delta}$ Single Crystals}
\author{P. Gyawali}
\author{V. Sandu\thanks}
\thanks{Permanent Address: National Institute of Materials Physics, 077125 Bucharest-Magurele, Romania}
\author{C. C. Almasan}
\affiliation{Department of Physics, Kent State University, Kent, Ohio 44242, USA}
\author{B. J. Taylor and M. B. Maple}
\affiliation{Department of Physics, University of California at San Diego, La Jolla, California 92903, USA}
\date{\today}
\begin{abstract}

Magnetic relaxation measurements on a series of Y$_{1-x}$Pr$_{x}$Ba$_{2}$Cu$_{3}$O$_{7-\delta}$ ($x = 0.13, 0.34, 0.47$)
single crystals were performed over a large field and temperature range in order to
investigate the characteristics of the vortex matter across the second magnetization peak (SMP).
The magnitude of the SMP varies non-monotonically with the Pr concentration; i.e.,  the irreversible magnetization 
normalized by its value at the onset field $H_{on}$ displays a maximum for the $x = 0.34$ single crystal. The two
characteristic fields, $H_{on}$ and $H_{sp}$, follow different temperature $T$ dependences:
$H_{on} \propto T^{\nu_{on}}$   and $H_{sp} \propto [1-(T/T_{c})^{2}]^{\nu_{sp}}$. The extracted values
of the apparent activation energy $U^\ast$ and the creep exponent $\mu$ display a maximum at a field
$H_{on} < H^\ast < H_{sp}$. Their field dependences point toward the coexistence of both elastic and
plastic creep for $H > H_{on}$. The degree of participation of each creep mechanism is determined by the charge carrier
density, which controls both the elastic properties of the vortex matter and the pinning potential.
\end{abstract}

\maketitle

\subsection{Introduction}
One of the most challenging manifestations of the vortex matter in cuprate superconductors is the
second magnetization peak (SMP) or fishtail displayed in the isothermal magnetization at an
intermediate magnetic field $H_{sp}$ much lower than the irreversibility field. \cite{Zhukov, Abulafia,
Giller, Jirsa, Reissner, Küpfer, Manson, Kokkaliaris, Avraham, Sun, Miu1} Theoretically, it has been shown that at this 
field
the flux line lattice crosses over from a dislocation-free Bragg glass structure present at low magnetic fields, which has algebraic decay
of translational correlations, to a vortex glass. \cite{Ertas, Vinokur,
Koshelev} The existence of this order-disorder transition was later observed experimentally. \cite{Nishizaki, Giller, 
Shibata} The magnetic field increases the entropy of the flux line system \cite{Watanabe} beyond that
introduced by the quenched disorder. Additionally, the field triggers the competition between the
elastic $U_{el}$ and pinning $U_{pin}$ energies at constant temperature (see, for example, Ref. 18). 
At low fields, the weak
point disorder is important and the flux lines form an elastic quasi-ordered dislocation-free lattice. As the field 
increases, the elastic screening length, \cite {Larkin} hence the interlayer interaction,
decreases, weakening the elastic response of the lattice and facilitating the invasion of topological
defects (dislocations). The break of the long range order allows a better matching of the vortex
lines to the pinning potential and, consequently, enhances the irreversibility present in 
magnetization. This effect starts at the onset field $H_{on}$. Above this field, the vortex matter, assumed to be a lattice, displays a
granular structure, namely, a network of dislocations that separate grains of vortex
matter that still retain a higher elasticity. \cite{Moretti}  The dislocation lines obey a characteristic
dynamics including pinning and depinning processes.\cite{Kierfeld1}

Although such a behavior is universal, each class of cuprates shows a specific behavior in which the second
magnetization peak is more or less pronounced, depending on the relative values of the elastic and pinning
energies, which, in turn, along with the characteristic
lengths (elastic screening length $L_{0}$ and Larkin length) depend on the superfluid density $n_{s}$ (through the 
magnetic penetration length $\lambda$ because the effective mass is almost $n_s$ independent),\cite{Padilla}
anisotropy $\gamma$, and the disorder parameter $\delta _{dis}$ (Refs. 19,23-25).  Usually, in the absence of a special treatment  e.g., particle irradiation, weak disorder due to oxygen vacancies is always
present in cuprates. 
In YBa$_2$Cu$_3$O$_{7-\delta}$, charge underdoping is obtained by oxygen removal from the Cu(1)-O(1) chains. This procedure induces also disorder consisting in different sequences of full Cu-O(1) and empty Cu(1) chains and an increased occupation of O(5) sites.\cite {Jorgensen} A second consequence of the underdoping is the fast increase in the anisotropy. Thus, the three parameters $n_{s}$, $\gamma$, and $\delta _{dis}$ are interconnected in a complex way. This makes  it difficult to systematically study the effect of one of these material parameters on the evolution of the vortex matter from a quasi-ordered Bragg glass to a disordered amorphous or liquid state (see, for example, Ref. 27).

It is possible to reduce the number of the above mentioned parameters from three to two by applying a doping method 
that practically does not introduce additional effective disorder. This is the case of Pr substitution for Y in
YBa$_2$Cu$_3$O$_{7-\delta}$. The
density of free charge carriers is extremely sensitive to Pr even at optimal oxygenation due to  carrier localization in
the Fehrenbacher-Rice band as a result of Pr-O hybridization,\cite{Fehrenbacher, Liechtenstein} while the disorder created
by the Pr ions is less effective to pinning compared with oxygen. \cite {Lobo} It is remarkable that the  integrity of the
chains, hence the orthorhombicity, is preserved in this material, notwithstanding the  change in the charge carrier
density. So, the main source of the weak pinning, i.e. oxygen disorder, is maintained almost constant. The only effect of
Pr on the pinning parameter results from its dependence on the magnetic penetration length.\cite{Blatter,Radzyner}
Therefore, the Pr substitution for Y in YBa$_2$Cu$_3$O$_{7-\delta}$ facilitates the investigation of the effect of the
superfluid density on the evolution and crossovers of different regimes of the vortex matter.

In this paper, we explore
the evolution of the vortex matter state  in the temperature and field range where the
second magnetization peak SMP is present by studying the magnetization and magnetic relaxation of a series of
Y$_{1-x}$Pr$_x$Ba$_{2}$Cu$_{3}$O$_{7-\delta}$ single crystals. Our study has shown that the main ingredient that controls 
the evolution of the vortex matter  through
the different regimes is the charge carrier density. The SMP is
first enhanced and then suppressed as $x$ increases. The reason for this behavior is the softening of the elastic moduli,
which makes the vortex lattice less stable to defect invasion. 

\subsection{Experimental Details}
A series of Y$_{1-x}$Pr$_x$Ba$_{2}$Cu$_{3}$O$_{7-\delta}$ ($x = 0.13, 0.34$, and $ 0.47$) single crystals was
chosen to perform extensive magnetization measurements. The three single crystals for which data are presented
here are platelets of size 1.25 x 0.95 x 0.12 mm$^{3}$, 0.75 x 0.4 x 0.1 mm$^{3}$, and 0.83 x 0.47 x 0.03
mm$^{3}$ with critical temperatures $T_{c}$ of 82, 50, and 34 K, respectively. Details of the
crystal growth are reported elsewhere.\cite{Paulius1}

Magnetic field $H$ dependent magnetization $M$ and relaxation measurements were performed at different temperatures
$T$ in the reduced temperature $T/T_c$ range between 0.2 and 0.9 by using a Quantum Design superconducting
quantum interference device magnetometer with the external magnetic field applied parallel to the $c$-axis of the
single crystal. We used the persistent current mode with a scan length of 40 mm, which guarantees an excellent magnetic
field homogeneity. The single crystals were cooled in zero field to the desired temperature and the whole $M(H)$ loop was
recorded in increasing and decreasing fields with  $H$ steps chosen to get the finest details in $M(H)$. After performing a
hysteresis loop at a given temperature, the sample was warmed up to $T >> T_{c}$ and zero-field-cooled to the next set
temperature.

For all the single crystals studied, the demagnetization factor $D$, calculated from the initial slope of the virgin
hysteresis curves, was found to be higher than 0.9, hence, the magnetic induction $B = H +4\pi(1- D)M \approx H$. Therefore, 
we have used the magnetic field $H$ throughout this paper instead of the magnetic induction $B$.

Magnetic relaxation was measured in the normal $dc$ mode by monitoring the time decay of the magnetic moment. For
these measurements, we zero-field-cooled the single crystal to the desired temperature, the magnetic field was then ramped
to the target value, and the magnetization was recorded as a function of time $t$ every 175 sec for about two hours. 

\subsection{Results and Discussion}

Figure 1 shows the magnetization loops $M(H)$ for three Pr concentrations measured at
$T/T_{c}\approx$ 0.3. These $M(H)$ curves are almost symmetric around the $M = 0$ axis, i.e., the reversible
magnetization is nearly zero, and they display a similar field dependence. Specifically, the
absolute value of the magnetization decreases with increasing field beyond the full penetration field,
reaches a minimum at the onset field $H_{on}$, then it increases again reaching a second magnetization peak (SMP) at
$H_{sp}$, and, finally, decreases to zero with further increasing the magnetic field. Previously,  both
$H_{on}$ and $H_{sp}$ have been related with the order-disorder transition from the dislocation-free Bragg glass to the 
disordered vortex matter with a glassy structure.
For example, several theoretical studies take the peak field $H_{sp}$ as the order-disorder line, \cite{Ertas, Vinokur,
Mikitik, Kierfeld2} while in experimental
studies of YBa$_2$Cu$_3$O$_{7-\delta}$, the order-disorder crossover has been taken either as the kink observed for $H_{on}
< H < H_{sp}$ (Refs. 15,32) or at $H_{on}$ (Refs. 3,33). 

The goal in the present work is to study
the effect of the charge carrier density $n_s$ on the evolution of the vortex matter in the temperature and field range
where the second magnetization peak SMP is present. A specific vortex state at a fixed $T$ and $H$ is determined by the
$n_s$-dependent interplay between the pinning landscape and the parameters that govern the  stability
of the vortex lattice, namely, disorder, critical current density $J_c$, and elastic moduli. The study of Pr-doped
YBa$_2$Cu$_3$O$_{7-\delta}$ insures that Pr doping does not change the disorder but only decreases $n_s$. Hence, by studying this system, one studies the above mentioned interplay as a function of $n_s$ only.
The decrease of
$J_c$ with decreasing
$n_s$ is reflected in the continuous decrease of the irreversible magnetization, including its value at the SMP, with
decreasing
$n_s$, i.e., increasing
$x$ (see Fig. 1). Hence, a plot of $M_{irr}/M_{irr,on}$, the irreversible
magnetization $M_{irr} \equiv( M^{-} - M^{+})/2$ ($M^-$ and $M^+$ are the magnetizations measured in decreasing and
increasing magnetic field, respectively) normalized by the value of the magnetization at the onset field $M_{irr,on}$,  
vs $H/H_{sp}$ eliminates the effect of $J_c(n_s)$. Therefore, such a plot reflects, at constant $H$, only the
$n_s$-dependent interplay between the pinning landscape and the elastic moduli, while, at constant doping $x$, it
refrects the interplay between the pinning landscape and the $H$-induced disorder.

A plot of $M_{irr}/M_{irr,on}$  
vs $H/H_{sp}$ measured at a reduced temperature $T/T_c=0.3$ and three different Pr dopings is given in
Fig. 2. Note that   the normalized irreversible magnetization in the SMP region displays a conspicuous
nonmonotonic behavior as a function of the charge carrier density, i.e., it increases with decreasing charge carrier
density down to $n_s=0.165$
($x = 0.34$) followed by a decrease for even lower $n_s$ (higher Pr concentrations). In fact, our magnetic investigations of
the $x = 0.53$ single crystals have shown that the SMP is absent at this Pr concentration. \cite{Sandu} 

The non-monotonic behavior of $M_{irr}/M_{irr,on}$  vs $H/H_{sp}$ can be explained as follows.  The elastic moduli
$C_{44}$ and
$C_{66}$ decrease with decreasing charge carrier density $n_s$ (increasing $x$).\cite{Blatter} The
decrease of the elastic moduli reduces the energy scale of the dislocations
\cite {Kierfeld1} [$E_D= (C_{44}C_{66})^{1/2}b/4\pi$, where $b$ is the Burger's vector], making easier their generation and,
subsequently, a better matching of the flux-line system to the $n_s$ dependent pinning landscape. Therefore, at high
charge carrier densities (e.g., the $x \sim 0.13$ single crystal), the flux-lines system is rather stiff since the creation
of dislocations requires a rather high energy. Hence, the density of dislocations is rather low and their relative
contribution to the total irreversibility, i.e., pinning of the flux lines, though important, is not substantial. At low
charge carrier densities (e.g., the $x \sim 0.47$ single crystal), the flux-lines system is soft since dislocations are
easily created ($C_{44}$ and $C_{66}$ are small, hence $E_D$ is low). Nevertheless, despite the increased
plasticity of the flux-lines system due to the increased number of dislocations, the values of $M_{irr}/M_{irr,on}$ are
smaller than the ones for the low Pr doping (high
$n_s$) case due to the net decrease in the strength of the
$n_s$-dependent pinning landscape. However, there is an optimum charge carrier density (e.g., the $x \sim 0.34$ single
crystal) for which the softening of the flux lattice due to the decrease in the elastic moduli allows the optimum
matching of the flux-lines system to the pinning landscape, maximazing the irreversible magnetization $M_{irr}/M_{irr,on}$.

The $H$ dependence of $M_{irr}/M_{irr,on}$ for a constant $x$ displays a broad maximum around the field $H_{sp}$
corresponding to the second magnetization peak. This sugests that the gradual process of field driven disordering gives
rise to  a continuous increase in the density of  dislocation loops. Hence, elastic and plastic behaviors coexist in
different degrees over a large field range  starting at $H_{on}$ (below which only elastic creep is present, Ref. 3) and ending far above
$H_{sp}$, with the elastic behavior dominant for $H<H_{sp}$ and plastic behavior dominant for $H>H_{sp}$. This
result is supported by STM investigations,\cite{Shibata} which show the presence of
dislocations even near $H_{on}$ and by neutron scattering data, which have shown surviving Bragg peaks far above $H_p$. (Ref. 35). Note that numerical simulations have shown that homogeneous flux 
line domains, i.e., systems displaying elastic creep, survive up to $0.8\times H_{c2}$ (Ref. 36).  

The overall behavior discussed above is present over the whole temperature range $0.3 \leq T/T_{c} \leq 0.9$. As
the temperature increases, both characteristic fields $H_{on}$ and $H_{sp}$ shift to lower values, but
they follow different $T$ dependences; namely, $H_{on} \propto T^{-\nu_{on}}$ (see Inset to Fig. 3) while  $H_{sp}
\propto [1-(T/T_{c})^{2}]^{\nu_{sp}}$  (see Fig. 3), with
$\nu_{on}=$  1.9, 2.1, and 2.2, and  $\nu_{sp}=$ 2.07, 2.53, and 2.23 for the $x =$ 0.13, 0.34, and 0.47 single crystal,
respectively.  A power law dependence of $H_{on}(T)$ was also reported for YBa$_2$Cu$_3$O$_{7-\delta}$ 
single crystals but with lower values of the exponent $\nu_{on}$, namely between 1.09 and 1.3 (Refs. 32,33).

The $H_{sp}(T)$ of the present study does not show the upturn at high temperatures reported for optimally doped
YBa$_2$Cu$_3$O$_{7-\delta}$ (Refs. 6,8,15,37). Nevertheless, previous measurements
have shown that the $H_{sp}(T)$ behavior is extremely sensitive to small deviations from optimal doping\cite{Küpfer,
Deligiannis} and that the suppression of the high temperature upturn for YBa$_2$Cu$_3$O$_{7-\delta}$ occurs for $\delta
\geq 0.06$. On the other hand, disorder introduced by electron irradiation was found to decrease, but not suppress, the
high temperature upturn of $H_{sp}(T)$ even for a fluence as high as
$2\times10^{18}$ cm$^{-2}$ (Ref.
38). Therefore, we conclude that the
reduction of the charge carrier density by Pr substitution for Y gives rise to the suppression of the upturn in
$H_{sp}(T)$ at high $T$, hence to the monotonic decrease of
$H_{sp}$ with increasing $T$. Specifically, the decrease in the
superfluid density $n_{s}$ with increasing doping $x$ at high $T$ increases the magnetic penetration length $\lambda$, hence decreases
$H_{sp}$ ($H_{sp} \propto 1/\lambda ^{4}$, Ref. 2). Notice that the
$x$ dependence of the exponent
$\nu_{sp}$ shows a similar nonmonotonic trend as the SMP itself. 

We also performed  magnetization relaxation measurements on
Y$_{1-x}$Pr$_x$Ba$_2$Cu$_3$O$_{7-\delta}$ at different fields and temperatures around the SMP in order to
further study the evolution of the vortex matter with the charge carrier density.  Figure 4 is a plot of the time $t$ dependence
of the irreversible magnetization $M_{irr}$  normalized by the first measured magnetization 
$M_{irr}(t_b)$ ($t_b \sim 250$ sec) for the 
$x = 0.34$ single crystal measured at a reduced temperature $T/T_{c} = 0.4$ and
different reduced magnetic fields $H/H_{sp}$ both below and above $H_{sp}$. We obtained $M_{irr}(t)$ by
subtracting the reversible magnetization, extracted from the $M(H)$ loops (see Fig. 1), from the measured magnetization.
The data of Fig. 4 are representative for all the single crystals measured. It is salient that the data do not follow a
logarithmic
$t$ dependence. Hence, we analyzed these data in the framework of the collective creep theory in which
\cite{Yeshurun}
\begin{equation}
M_{irr}(t,T,H) = M_{irr}(t_0,T,H)\biggr[1+\frac{\mu k_{B}T}{U_{0}(H)}
\ln\biggr(\frac{t}{t_{0}}\biggr)\biggr]^{-\frac{1}{\mu}},
\end{equation}
where $U_{0}$ is the effective pinning
potential,
$t_{0}$ is a macroscopic quantity depending on the sample size and it should not be confused with the actual microscopic
attempt time,\cite {Yeshurun} and
$\mu$ is the collective creep exponent. The normalized relaxation rate $S$ is obtained from Eq. (1) as
\begin{eqnarray}
S(J,T,H)\equiv - \frac{1}{M_{irr}(t)}\frac{dM_{irr}(t)}{d\ln(t)} = \frac{k_{B}T}{U^{\ast}(J,T,H)},
\end{eqnarray} 
where
\begin{equation}
U^\ast(J,T,H) \equiv U_{0}+ \mu k_{B}T \ln\biggr(\frac{t}{t_{0}}\biggr),
\end{equation} 
is the apparent activation energy, which is larger than $U_0$ due to current relaxation. Note
that, for convinience, we define $S$ as a positive quantity. One can determine experimentally $U^{\ast}$ from Eq. (2) 
with the normalized relaxation rate $S$ obtained from the data of Fig. 4. In doing so, the relaxation rate is normalized to the
initial magnetization $M_{irr}(t_b)$ rather than the time dependent magnetization. Since the variation in $M_{irr}$ during the
relaxation measurement is small, the error introduced is also small.\cite{Yeshurun} Before discussing the physics that $U^{\ast}$
would reveal, we discuss next the relationship between this apparent activation energy, which is accessible experimentally, and the
actual activation energy.

The actual activation energy $U$ is a rather complex quantity involving not only a term due to the microscopic
interaction between the flux lines and the pinning centers $U_{int}$, but also an extrinsic dependence on the distribution
of the critical current density.\cite{JSun}  Miu et al.,\cite{Miu2} inferred that  
\begin{equation}
U =
U_{int}(J)\ln(J_{c}/J). 
\end{equation}
Then, the relationship between $U^\ast$ and $U$, hence
$U_{int}$, can be obtained through the general dependence
$U = k_{B}T\ln(t/t_{0})$ (Ref. 42) along with Eqs. (2) and (4) (see the Apendix for more details) as:
\begin{equation}
U^\ast(J,H) = U_{int}(J) - J \frac{dU_{int}}{dJ}\ln\biggr(\frac{J_{c}}{J}\biggr).
\end{equation}
Therefore, $U^\ast$  
is always an overestimate of $U_{int}$ since $dU_{int}/dJ$ is negative.
Nevertheless, Eq. (5) shows that the approximation of $U_{int}$ with
$U^\ast$ is valid as long as the current density is close to $J_c$ and it breaks down for $J \ll J_c$ i.e., for $H\gg
H_{sp}$.
Thus, a plot of  $U^\ast(J)$ determined from Eq. (2) with $S$ obtained from the relaxation data in the
regime where
$J$ is not too far from
$J_c$, i.e.,  for not extremely long relaxation times and for magnetic fields around $H_{sp}$, gives an accurate information
on
$U_{int}$, hence, on the evolution of the vortex matter when the temperature and magnetic field are swept. 

The plot of $U^\ast$ vs $M_{irr}$ ($M_{irr}
\propto J$) is shown in Fig. 5 for the $x = 0.13$ single
crystal measured at
$T/T_{c} \approx 0.4$ and different values of the reduced field $H/H_{sp}$. It is salient the different evolution of the
activation energy below and above $H_{sp}$. Specifically, for fields smaller than $H_{sp}$ (open symbols),
$U^\ast(J,H)$ increases rapidly as the current $J$ (or equivalently $M_{irr}$) decreases, an expected behavior for an
elastic vortex system in the collective pinning regime. Above $H_{sp}$ (filled symbols), the increase of
$U^\ast$ with decreasing current density becomes slower and slower suggesting a smooth crossover at $H_{sp}$ to another
regime, most likely a regime dominated by the fast proliferation of dislocations. 

We have observed a similar behavior of $U^\ast(J)$ for  all Pr concentrations studied.
However, for the same reduced field range and for $T/T_{c} \approx 0.4$, the range of values of $U^\ast$
decreases with increasing Pr concentration; e.g.,
$U^\ast$ varies between 200 - 1000 K for the $x = 0.13$, 100  - 600 K for the $x = 0.34$, and 50 - 400 K for
the $x = 0.47$ single crystal. The decrease in the range of
$U^\ast$ with increasing $x$ is in agreement with previous results on
Y$_{1-x}$Pr$_{x}$Ba$_{2}$Cu$_{3}$O$_{7-\delta}$ polycrystalline pellets.
\cite{Paulius2} 

As a function of field, $U^\ast$ (or equivalently S, see Eq. (2)) systematically shows, at any relaxation time $t$, a
maximum (minimum) at a field $H^\ast$, which is between 
$H_{on}$ and
$H_{sp}$ [see, for example, the Inset to Fig. 5 for $S(H)$ for the $x = 0.34$ single
crystal]. Therefore, the slowest magnetic relaxation takes place at a field value just below $H_{sp}$ and not at $H_{sp}$.
At fields higher than
$H^\ast$,
$S(H)$ increases almost linearly.

The apparent activation energy $U^\ast$ is a nonlinear function of time, but a good estimate of its field dependence in a certain time window $t_1$ to $t_2$ can be obtained by replacing the derivative in Eq. (2) with finite differences.  This average apparent activation energy 
$\overline{U}^\ast$ is given by 
\begin{equation}
\space \overline{U}^\ast(H) = k_{B}T\ln(\frac{t_{2}}{t_{1}})\frac{M_{irr}(t_b,H)}{\triangle M_{irr}(H)},
\end{equation}
where $\triangle M_{irr}(H) = M_{irr}(H, t_{1}) - M_{irr}(H, t_{2})$.
Figure 6 is a log-log plot of $\overline{U}^\ast(H)$, as extracted from relaxation measurements over the time
window 600 sec $\leq t \leq$ 4000 sec, for three Pr concentrations ($x$ = 0.13, 0.34, and 0.47). $\overline{U}^\ast(H)$  
displays again a maximum at $H^{\ast}$.  
The relative difference $(H_{sp}-H^{\ast})/H_{sp}$ between
$H^\ast$ and $H_{sp}$ is maximum for the $x = 0.34$ single  crystal, for which the value of the second magnetization
peak is enhanced (see Fig. 2). For $H < H^\ast$, $\overline{U}^\ast$  increases with increasing magnetic field, which is
consistent with the elastic (collective) creep mechanism. For $H>H^\ast$,
$\overline{U}^\ast(H)$    decreases roughly as a power law with increasing $H$, i.e., $\overline{U}^\ast(H)
\propto H^{-\nu}$, with $\nu$ = 0.6 and 0.4 for the $x = 0.34$ and $0.47$ single crystal, respectively,
which indicates plastic vortex creep,
\cite{Giller} and a more abrupt drop for the $x = 0.13$ single crystal. A simple model for plastic pinning
\cite{Abulafia} yields an exponent $\nu$ = 0.5, while other reports give $\nu = 0.55$ for
Bi$_{2}$Sr$_{2}$CaCu$_{2}$O$_{8}$ (Ref. 44),
$\nu = 0.7$ for YBa$_{2}$Cu$_{3}$O$_{7-\delta}$ (Ref. 2) and HgBa$_{2}$CuO$_{4+\delta}$ (Ref. 45), 
and
$\nu\approx  0.9$ for Tl$_{2}$Ba$_{2}$CaCu$_{2}$O$_{8}$ (Ref. 46). Hence, as expected, these
$\overline{U}^\ast(H)$ data are consistent with the $U^{\ast}(J,H)$ data of Fig. 5, but, additionally, they give the
quantitative dependence of the activation energy on the magnetic field. Based on these data, we also conclude that, at least
in the case of  Pr-doped YBa$_2$Cu$_3$O$_{7-\delta}$, the order-disorder crossover is best given by $H^{\ast}(T)$ ($H_{on}<H<H_{sp}$),
eventhough, as discussed above, the two characteristic fields $H_{on}$ and $H_{sp}$ are the ones which previously have been
related with the order-disorder transition. 

Based on our above conclusion,  leveling the elastic and plastic energies engaged 
in the equilibrium of the flux lines,  i.e.,  $U_{el} = U_{pl}$, at $H^{\ast}$ gives $H^{\ast} \propto 1/ T^2 \lambda^4 \propto [\{1- (T/T_{c})^4\}/T]^{\chi}$, with $\chi=2$. The fit of $H^{\ast}$ vs $[1- (T/T_{c})^4]/T$ for $0.3\leq T/T_{c}\leq 0.8$ is shown as a log - log plot in the inset to Fig.  6. The values of the exponent $\chi$ are $1.44, 1.27$, and $1.54$ for the $x = 0.13, 0.34$, and $0.47$ single crystal, respectively. The good fit of the data with the above expression supports our conclusion that the elastic to plastic crossover takes place at $H^{\ast}$.
It is interesting to note that the exponents are non-monotonic with increasing $x$, with the lowest value for the $x = 0.34$ sample. 

Since, as shown above, the activation energy is a function of current density, temperature, and magnetic field [see Eq.
(3)], changes of any one of these parameters drive continuously the vortex matter into different elastic and plastic
creep regimes. Hence, magnetic relaxation data give information about a specific
flux-creep regime for a given $T$ and $H$ through the critical exponent
$\mu(T,H)$ present in Eq. (1). A fit of the $M_{irr}$ vs $\ln t$ data with Eq. (1) for different $T$ and $H$ gives
$\mu(T,H)$. 

Figure 7 and its insets are plots of $\mu(T,H)$ for different charge carrier densities, i.e., Pr doping. Note that
$\mu(H)$ displays a peak at the same magnetic field value $H^{\ast}$ at which $\overline{U}^\ast(H)$ is maximum. 
Therefore, as discussed above, that elastic pinning mechanism dominates for $H_{on} <H<H^{\ast}$ while the plastic
mechanism dominates for
$H>H^{\ast}$.  Also, note
that $\mu$ decreases with increasing temperature and decreasing charge carrier density  (increasing $x$). 

The collective
(elastic) creep theory\cite{Blatter} predicts that
$\mu = 1/7$  for single vortex creep (at high current and low field),
$\mu=5/2$ for small vortex-bundles creep (at intermediate current and field), $\mu=1$ for the creep of intermediate
vortex bundles, and
$\mu=7/9$ for the creep of large vortex bundles (at low current and high field). As the data show, $\mu (H)$ does not follow exactly these theoretical predictions in the regime where flux lines are expected to behave elastically, i.e., for $H_{on} <H < H^{\ast}$. For example, $\mu > 2$ but smaller than 2.5 for the lowest measured temperature ($T/T_c=0.3$) even in the case of the $x = 0.13$ single crystal, which is expected to have the strongest elastic response. Hence, although these values indicate that the relaxation of the flux vortices is mainly due to the creep of small vortex bundles, the admixture of the plastic contribution limits $\mu$ to values smaller than 2.5. Similar deviations from the theoretical exponents for plastic creep \cite{Kierfeld1} are clear for $H > H^{\ast}$.

The direct (inverse) correlation
between the values of the creep exponent
$\mu$ and the the charge carrier density (Pr doping)  as well as the the decrease of the value of $\mu$ with increasing $T$ (see Fig. 7 and its insets) reflects the decreased role of the
vortex lattice properties over single vortex behavior with decreasing $n_s$ or increasing $T$ as a result of the  weakening of the
elastic moduli with decreasing 
$n_{s}$ [$C_{66}\propto \lambda^{-2}\propto   n_{s}(x)$] or increasing $T$. As a consequence, the crossover magnetic field $H_{sb}$ from
single vortex  to small bundle collective pinning increases with decreasing $n_{s}$ (increasing $x$) or increasing $T$ since  $H_{sb} \propto
\lambda^{8/3}\propto n_{s}^{-4/3}$ (Ref. 23). Hence, $H_{sb}$ cannot be reached in strongly under-doped single crystals before the full crossover to plastic
pinning. This explains the decrease of the value of the exponent $\mu$ at
$H^{\ast}$ in Fig. 7 and its insets with increasing $x$ or $T$. 

For magnetic fields higher than
$H^{\ast}$,
$\mu$ decreases monotonically with increasing $H$. However, the value of $\mu=10/21$, representative for the plastic
creep of the lattice, is accessible only at high temperatures. At lower $T$ one needs magnetic fields higher than the one
available (5 T) in order to be able to detect this regime. Additionally, the elastic contributions do not vanish
completely in the plastic regime.\cite{Chandran}

\subsection{Summary} 
In summary, we investigated the evolution of the second magnetization peak (SMP) with the charge carrier
density by an appropriate doping that avoids the change of the quenched disorder in a sensitive way.
For this goal, we carried out magnetization and magnetic relaxation measurements on a series of
Y$_{1-x}$Pr$_{x}$Ba$_{2}$Cu$_{3}$O$_{7-\delta}$ single crystals in which the concentration of Pr ions controls
the charge carrier density $n_{s}(x)$.  We have found that the
quenched disorder is necessary for the existence of irreversibility and of the SMP, but the principal ingredient which
controls the evolution of the vortex matter through different regimes is the charge carrier density. Specifically, we have
found that the SMP is broad and its magnitude is non monotonic with the amount of doping: it increases with decreasing the
charge carrier density up to a doping around
$x=0.34$ followed by a decrease with further decreasing $n_s$ (increasing Pr concentration). The two characteristic
magnetic fields, the onset field
$H_{on}$ and the field
$H_{sp}$ corresponding to the SMP decrease with increasing temperature $T$, but they follow different $T$
dependences: $H_{on} \propto T^{\nu_{on}}$ while  $H_{sp}\propto [1-(T/T_{c})^{2}]^{\nu_{sp}}$, with the exponent $\nu_{sp}$
following the same nonmonotonic trend as a function of $n_s$ as
$H_{sp}$. Within the collective creep theory, we determined the apparent
activation energy. Its evolution with
$J$ has shown that the vortex system is predominantly elastically pinned below $H_{sp}$, while above  $H_{sp}$
there is a smooth crossover to a vortex regime most likely dominated by the proliferation of dislocations. The field
dependence of the average apparent pinning potential $\overline{U}^\ast$ displays a maximum at a magnetic field $H^\ast$,
with $H_{on} < H^\ast < H_{sp}$, which is consistent with the presence of an elastic (collective) creep mechanism at low
fields and plastic vortex creep at high $H$ values. The transition from the Bragg glass to the dislocation rich vortex system 
occurs gradually and extends on a rather large field range. For this reason, we propose that the order-disorder line must
be defined by the maximum of the average activation energy, which is located at $H^{\ast}$, below $H_{sp}$ but above the inflection point
of the $M(H)$ curves.
\\
\\
\textbf{Acknowledgments}
This research was supported by the National Science
Foundation under Grant No. DMR-0705959 at KSU, the US Department of Energy under Grant
No. DE-FG02-04ER46105 at UCSD, and NASR under Grant CEX 45/2006 at NIMP. \label{}

\subsection{Appendix}

As mentioned in the main text, Miu et al,\cite{Miu2} inferred the following current dependence of the activation energy:
\begin{displaymath}
U(J)=U_{int}(J) \ln(\frac{J_{c}}{J}).
\end{displaymath}
Then,
\begin{displaymath}
\frac{dU}{dJ} =  \frac{dU_{int}}{dJ}\ln\Big(\frac{J_{c}}{J}\Big) + U_{int} \frac{d(\ln J_{c} - \ln J)}{dJ}\\  
\end{displaymath}
\begin{displaymath}
  =  \frac{dU_{int}}{dJ} \ln\Big(\frac{J_{c}}{J}\Big) - \frac{U_{int}}{J},\hspace{0.5 in}
\end{displaymath}
or
\begin{equation}
-J \Big( \frac{dU}{dJ}\Big) =U_{int} - J \frac{dU_{int}}{dJ}\ln\Big(\frac{J_{c}}{J}\Big).
\end{equation}
As shown by Eq. (2), the normalized relaxation rate is
\begin{equation}
- \frac{1}{M_{irr}(t)}\frac{dM_{irr}(t)}{d\ln(t)} = \frac{k_{B}T}{U^{\ast}(J,T,H)}.
\end{equation}
Also, the actual activation energy  given by
\begin{displaymath}
U= k_{B}T\ln (t/t_{0})
\end{displaymath}
implies
\begin{equation}
\frac{dU}{d\ln t} = k_{B}T.
\end{equation}
Since $M_{irr}\propto J$, Eq. (8) becomes
\begin{displaymath}
-\frac{1}{J}\frac{dJ}{d\ln t} =\frac{dU}{d\ln t} \frac{1}{U^\ast},
\end{displaymath}
hence, 
\begin{equation}
U^\ast = -J\frac{dU}{dJ}.
\end{equation}
Using Eq. (7), the above equation gives 
\begin{displaymath}
U^\ast = -J\frac{dU}{dI} = U_{int} - J \frac{dU_{int}}{dJ}\ln\Big(\frac{J_{c}}{J}\Big).
\end{displaymath}

\newpage
\section{Figure Captions}
Fig. 1 (Color online) Magnetic hysteresis loops for Y$_{1-x}$Pr$_{x}$Ba$_{2}$Cu$_{3}$O$_{7-\delta}$ ($x = 0.13, 0.34$ and
$0.47$) single crystals, measured at the same reduced temperature $T/T_{c} \approx 0.3$. The arrows
indicate the position of the onset $H_{on}$ and second magnetization peak $H_{sp}$ fields.

Fig. 2 (Color online)  Plot of the irreversible magnetization $M_{irr}$ normalized to its value $M_{irr,on}$ at
the onset of the second magnetization peak, as a function of the reduced field $H/H_{sp}$ for
Y$_{1-x}$Pr$_{x}$Ba$_{2}$Cu$_{3}$O$_{7-\delta}$ ($x = 0.13, 0.34$ and $0.47$) single crystals  measured at the same
reduced temperature T/$T_{c} \approx 0.3$. Inset: $x$ dependence of the absolute value of the irreversible magnetization at
the second magnetization peak $M_{irr, sp}$, measured at the same reduced temperature.

Fig. 3 (Color online) Log-log plot of the second magnetization peak $H_{sp}$ vs reduced temperature $T/T_c$ for
Y$_{1-x}$Pr$_{x}$Ba$_{2}$Cu$_{3}$O$_{7-\delta}$ ($x = 0.13, 0.34$ and $0.47$) single crystals. Inset:  Log-log plot
of the onset field $H_{on}$ vs $T/T_c$ for the same single crystals.

Fig. 4 (Color online) Logarithm of time $t$ evolution of the irreversible magnetization $M_{irr}$ of an 
Y$_{0.66}$Pr$_{0.34}$Ba$_{2}$Cu$_{3}$O$_{7-\delta}$ single crystal measured  at the reduced temperature $T/T_{c}$ = 0.4 and different field values around the second magnetization peak. Open symbols below $H_{sp}$, and partially filled or close symbols above $H_{sp}$ represents 1.06, 1.21, 1.51, 1.82, 2.12, 2.42, 2.73, 2.03, 3.64 as shown by arrow.  

Fig. 5 (Color online) Plot of the apparent activation energy $U^\ast$, obtained from relaxation
measurements, vs irreversible magnetization $M_{irr}$ of an Y$_{0.87}$Pr$_{0.13}$Ba$_{2}$Cu$_{3}$O$_{7-\delta}$ single
crystal measured at different applied magnetic fields and at the reduced temperature $T/T_{c}\approx 0.4$. Empty and filled symbols are for $H< H_{sp}$ and$H> H_{sp}$, respectively. Solid lines are guide for the eye. Inset: Field dependence of the relaxation rate $S$ measured 
at a time $t \approx 6000 sec$.

Fig. 6 (Color online) Log-log plot of the field $H$ dependence of the average activation energy $\overline U^\ast$   of
Y$_{1-x}$Pr$_{x}$Ba$_{2}$Cu$_{3}$O$_{7-\delta}$ ($x = 0.13, 0.34$ and $0.47$) single crystals measured at a reduced temperature $T/T_c=0.4$. The lines are fits of the data  with a power law. Inset: Temperature dependence of $H^{\ast}(T)$ determined from the minima of $S(H)$.

Fig. 7 (Color online) Field $H$ dependence of the relaxation exponent $\mu$ of an
Y$_{0.87}$Pr$_{0.13}$Ba$_{2}$Cu$_{3}$O$_{7-\delta}$ single crystal measured at different reduced temperatures ($T_{c}$ =
82 K). Insets: The same plot for the (a) Y$_{0.66}$Pr$_{0.34}$Ba$_{2}$Cu$_{3}$O$_{7-\delta}$ single crystal ($T_{c}$ =
50 K)  and (b) Y$_{0.53}$Pr$_{0.47}$Ba$_{2}$Cu$_{3}$O$_{7-\delta}$ single crystal ($T_{c}$ = 34 K).


\begin{thebibliography}{99}
\bibitem{Zhukov}A. A. Zhukov, H. K\"upfer, G. Perkins, L. F. Cohen, A. D. Caplin, S. A. Klestov, H. Claus, V.I. 
Voronkova, T. Wolf, and H. W\"uhl, Phys. Rev. B \textbf{51}, 12 704 (1995).
\bibitem{Abulafia}Y. Abulafia, A. Shaulov, Y. Wolfus, R. Prozorov, L. Burlachkov, Y. Yeshurun, D. Majer, E.
Zeldov, H. W\"uhl, V. B. Geshkenbein, and V. M. Vinokur, Phys. Rev. Lett.\textbf{77}, 1596 (1996).
\bibitem{Giller}D. Giller, A. Shaulov, R. Prozorov, Y. Abulafia, Y. Wolfus, L. Burlachkov, Y. Yeshurun, E.
Zeldov, V. M. Vinokur, J. L. Peng, and R. L. Greene, Phys. Rev. Lett.\textbf{79}, 2542 (1997).
\bibitem{Jirsa}M. Jirsa, L. P\r{u}st, D. Dlouh\'y, and M. R. Koblischka, Phys. Rev. B \textbf{55}, 3276 (1997)
\bibitem{Reissner}M. Reissner and J. Lorenz, Phys. Rev. B \textbf{56}, 6273 (1997).
\bibitem{Küpfer}H. K\"upfer, Th. Wolf, C. Lessing, A. A. Zhukov, X. Lan\c con, R. Meier-Hirmer, W. Schauer, and
H. W\"uhl, Phys. Rev. B \textbf{58}, 2886 (1998).
\bibitem{Manson}J. T. Manson, J. Giapintzakis, and D. M. Ginsberg, Phys. Rev. B \textbf{54}, 12 517 (1997).
\bibitem{Kokkaliaris}S. Kokkaliaris, P. A. J. de Groot, S. N. Gordeev, A. A. Zhukov, R. Gagnon, and L.
Taillefer, Phys. Rev. Lett. \textbf{82}, 5116 (1999).
\bibitem{Avraham}N. Avraham, B. Khaykovich, Y. Myasoedov, M. Rappaport, H. Shtrikman, D. E. Feldman, T.
Tamegai, P. H. Kes, M.  Li, M. Konczykowski, K. V. D. Beek, and E. Zeldov,  Nature  \textbf{411}, 451 (2001).
\bibitem{Sun}Y. P. Sun, W. H. Song, J. J. Du, and H. C. Ku, Phys. Rev. B \textbf{66}, 104520 (2002).
\bibitem{Miu1}L. Miu, S. Popa, T. Noji, Y. Koike, D. Miu, S. Diaz, and G. Chouteau, Phys. Rev. B \textbf{70},134523 (2004).
\bibitem{Ertas}D. Ertas and D. R. Nelson, Physica C \textbf{272}, 79 (1996).
\bibitem{Vinokur}V. Vinokur, B. Khaykovich, E. Zeldov, M. Konczykowski, R. A. Doyle, and P. H. Kes, Physica
C \textbf{295}, 209 (1998).
\bibitem{Koshelev}A. E. Koshelev and V. M. Vinokur, Phys. Rev. B \textbf{57}, 8026 (1998).
\bibitem{Nishizaki}T. Nishizaki, T. Naito, and N. Kobayashi, Phys. Rev. B \textbf{58}, 11169 (1998)
\bibitem{Shibata}K. Shibata, T. Nishizaki, M. Maki, and N. Kobayashi, Phys. Rev B \textbf{72}, 014525 (2005).
\bibitem{Watanabe}K. Watanabe, T. Kita, and M. Arai, Phys. Rev. B \textbf{71}, 144515 (2005).
\bibitem{Radzyner}Y. Radzyner, A. Shaulov, and Y. Yeshurun, Phys. Rev. B \textbf{65}, 100513 (2002).
\bibitem{Larkin}A. I. Larkin and V. M. Vinokur, Phys. Rev. Lett. \textbf{75}, 4666 (1995).
\bibitem{Moretti}P. Moretti, M.-C. Miguel, M. Zaiser, and S. Zapperi, Phys. Rev. Lett. \textbf{92}, 257004 (2004).
\bibitem{Kierfeld1}J. Kierfeld, H. Nordborg, and V. M. Vinokur, Phys. Rev. Lett. \textbf{85}, 4948 (2000).
\bibitem{Padilla}W. J. Padilla, Y. S. Lee, M. Dumm, G. Blumberg, S. Ono, K. Segawa, S. Komiya, Y. Ando, and D. N. Basov, Phys. Rev. B \textbf{72}, 060511(R) (2005).
\bibitem{Blatter}G. Blatter, M. V. Feigel'man, V. B. Geshkenbein, A. I. Larkin, and V. M. Vinokur, Rev. Mod.
Phys. \textbf{66}, 1125 (1994).
\bibitem{Mikitik}G. P. Mikitik and E. H. Brandt, Phys. Rev. B \textbf{64}, 184514 (2001)
\bibitem{Kierfeld2}J. Kierfeld and V. Vinokur, Phys.  Rev. B \textbf{69}, 024501 (2004).

\bibitem{Jorgensen}J. D. Jorgensen, B. W. Veal, A. P. Paulikas, L. J. Nowicki, G. W. Crabtree, H. Claus, and W. K. Kwok, Phys. Rev. B \textbf{41}, 1863 (1990).
\bibitem{Baziljevich}M. Baziljevich, D. Giller, M. McElfresh, Y. Abulafia, Y. Radzyner, J. Schneck, T. H.
Johansen, and Y. Yeshurun, Phys. Rev. B \textbf{62}, 4058 (2000).
\bibitem{Fehrenbacher} R. Fahrenbacherb and T. M. Rice, Phys. Rev. Lett. \textbf{70}, 3471 (1993)
\bibitem{Liechtenstein}A. I. Liechtenstein and I. I. Mazin, Phys. Rev. Lett. \textbf{74}, 1000 (1995).
\bibitem{Lobo}R. P. S. M. Lobo, E. Ya. Sherman, D. Racah, Y. Dagan, and N. Bontemps, Phys. Rev. B \textbf{65}, 104509 (2002).
\bibitem{Paulius1}L. M. Paulius, B. W. Lee, M. B. Maple, and P. K. Tsai, Physica C \textbf{230}, 255 (1994).
\bibitem{Pissas1}M. Pissas, E. Moraitakis, G. Kallias, and A. Bondarenko, Phys. Rev. B \textbf{62}, 1446 (2000).
\bibitem{Pal}D. Pal, S. Ramakrishnan, A. K. Grover, D. Dasgupta, and B. K. Sarma, Phys. Rev. B \textbf{63}, 132505 (2001).
\bibitem{Sandu}V. Sandu, P. Gyawali, T. Katuwal, and C. C. Almasan, B. J. Taylor and M. B. Maple, Phys. Rev. B \textbf{74}, 184511 (2006).
\bibitem{Pautrat} A. Pautrat, Ch. Simon, C. Goupil, P. Mathieu, A. Bržlet, C. D. Dewhurst, and A. I. Rykov, Phys. Rev. B \textbf{75}, 224512 (2007).
\bibitem{Chandran}M. Chandran, R. T. Scalettar, and G. T. Zim\'anyi, Phys. Rev. B \textbf{69}, 024526 (2004).
\bibitem{Deligiannis}K. Deligiannis, P. A. J. de Groot, M. Oussena, S. Pinfold, R. Langan, R. Gagnon, and L. Taillefer, Phys. Rev. Lett. \textbf{79}, 2121 (1997).
\bibitem{Nishizaki1}T. Nishizaki, T. Naito, S. Okayasu, A. Iwase, and N. Kobayashi, Phys. Rev. B \textbf{61}, 3649 (2000).
\bibitem{Yeshurun}Y. Yeshurun, A. P, Malozemoff, and A. Shaulov, Rev. Mod. Phys. \textbf{68}, 911 (1996).
\bibitem{JSun}J. Z. Sun, C. B. Eom, B. Lairson, J. C. Bravman, and T. H. Geballe, Phys. Rev. B \textbf{43}, 3002 (1991).
\bibitem{Miu2}L. Miu, T. Noji, Y. Koike, E. Cimpoiasu, T. Stein, and C. C. Almasan, Phys. Rev. B \textbf{62}, 15172 (2000).
\bibitem{Feigel'man}M. V. Feigel'man, V. B. Geshkenbein, A. I. Larkin, and V. M. Vinokur, Phys. Rev. Lett. \textbf{63}, 2303 (1989).
\bibitem{Paulius2}L. M. Paulius, C. C. Almasan, and M. B. Maple, Phys. Rev. B \textbf{47}, 11627 (1993).
\bibitem{Miu3} L. Miu, E. Cimpoiasu, T. Stein and C. C. Almasan,Physica C \textbf{334}, 1 (2000).
\bibitem{Pissas}M. Pissas, D. Stamopoulos, E. Moraitakis, G. Kallias, D. Niarchos, and M. Charalambous,
Phys. Rev. B \textbf{59}, 12121, (1999).
\bibitem{Chowdhury}P. Chowdhury, H.-J. Kim, In - Sun Jo, and S.-I. Lee,  Phys. Rev. B
\textbf{66}, 184509, (2002).

\end{thebibliography}
\end{document}